\documentclass[preprint,showpacs,preprintnumbers,amsmath,amssymb]{revtex4}
\usepackage{url}
\usepackage{helvet}
\usepackage{times}

\usepackage{dcolumn}
\usepackage{bm} 
\usepackage{amssymb}
\usepackage{graphicx}
\usepackage{amsfonts}
\usepackage{amsmath}
\usepackage{color}
\usepackage[applemac]{inputenc}

\begin{document}

\title{On negative higher-order Kerr effect and filamentation}

\author{V. Loriot$^{1,3}$}
\author{P. B\'ejot$^{{1,2}}$}
\author{W. Ettoumi$^{{2}}$}
\author{Y. Petit$^{2}$}
\author{J. Kasparian$^{2}$}\email{jerome.kasparian@unige.ch}
\author{S.~Henin$^{2}$}
\author{E. Hertz$^{1}$}
\author{B. Lavorel$^{{1}}$}
\author{O. Faucher$^{{1}}$}\email{olivier.faucher@u-bourgogne.fr}
\author{J.-P.~Wolf$^{{2}} $}


\affiliation{ Laboratoire Interdisciplinaire Carnot de Bourgogne (ICB), UMR 5209 CNRS-Universit\'e de Bourgogne, 9 Av. A. Savary, BP 47 870, F-21078 Dijon Cedex, France}
\affiliation{Universit\'e de Gen\`eve, GAP-Biophotonics, 20 rue de l'Ecole de M\'edecine, 1211 Geneva 4, Switzerland}
\affiliation{Present address: Instituto de Qu\'imica F\'isica Rocasolano, CSIC, C/Serrano, 119, 28006 Madrid, Spain}

\begin{abstract}
As a contribution to the ongoing controversy about the role of higher-order Kerr effect (HOKE) in laser filamentation, we first provide thorough details about the protocol that has been employed to infer  the HOKE indices from the experiment. Next, we discuss potential sources of artifact in the experimental measurements of these terms and show that neither the value of the observed birefringence, nor its inversion, nor the intensity at which it is observed, appear to be flawed.
Furthermore, we argue that, independently on our values, the principle of including HOKE is straightforward. Due to the different temporal and spectral dynamics,  the respective efficiency of defocusing by the plasma and by the HOKE is  expected to depend substantially on both incident wavelength and  pulse duration. The discussion should therefore focus on defining the conditions where each filamentation regime dominates.
\end{abstract}

\pacs{42.65.Jx Beam trapping, self focusing and defocusing,
self-phase modulation; 42.65.Tg Optical solitons; 78.20.Ci Optical, 37.10.Vz
Mechanical effects of light on atoms, molecules, and ions; 42.50.Md Optical transient phenomena: quantum beats, photon echo, free-induction decay, dephasings and revivals, optical nutation, and self-induced transparency
constants}

\maketitle

\section{Introduction}

While potentially spectacular applications like rainmaking or the control of lightning \cite{KaspaRMYSWBFAMSWW2003,KaspaAAMMPRSSYMSWW2008a,RohweKSHHLNPQSSWW2010a} attract much attention on the filamentation of ultrashort laser pulses, a controversy has recently been raised about the physical mechanism at the root of this phenomenon. Filaments are generally described as a dynamic balance between Kerr self-focusing and defocusing by the plasma generated at the non-linear focus \cite{ChinHLLTABKKS2005,BergeSNKW2007,CouaiM2007,KaspaW2008}. The recent measurement of higher-order Kerr effect (HOKE) indices of alternate signs in air and argon \cite{Loriot17_2009,Loriot18_2010} led us to propose that these terms provide the main defocusing effect, so that ionization and self-guiding are almost decoupled \cite{BejotKHLVHFLW2010a}, at least in the visible and infrared regions \cite{EttouBPLHFLKW2010}. Based on numerical work, and in the lack of knowledge of their values, these terms had already been assumed to contribute to defocusing, but only marginally \cite{AkoSBC2001,VincoB2004,FibicI2004,BergSMKYFSW2005,ZhangTWDW2010}. Our unexpected prediction has therefore been actively challenged \cite{KolesWM2010,PolynKWM2010,TelekWK2010}.


The controversy simultaneously focuses on two questions: the validity of the experimental measurement of the HOKE indices, and the validity of a filamentation model based on them. It is fed by the difficulty to perform quantitative measurements in filaments, due to the high intensity within them. This difficulty prevents one to directly test the contribution of the HOKE to filamentation. 
In this paper, we address these two aspects, with the aim of making the controversy as factual as possible by summarizing the facts and opening questions on this subject. In a first section, we establish the methodology used for extracting the HOKE terms from the experiment and discuss several potential artifacts in the experimental measurement of the HOKE terms, the values of which are critical to evaluate their contribution to filamentation. In a second section, we discuss the relevance and the physical consequences of the introduction of the HOKE in the description of filamentation. We suggest that the contribution of the HOKE on the filamentation process strongly depends on the incident wavelength and pulse duration. More specifically, longer wavelengths and shorter pulses are more sensitive to the HOKE, while defocusing by the plasma is favored in the case of shorter wavelengths and long pulses.

\section{On the measurement of the HOKE indices}

The key result reported by Loriot \textit{et al.}\,\cite{Loriot17_2009,Loriot18_2010} is the saturation and inversion of the instantaneous (i.e., at least, much shorter than the experimental resolution of $\sim$100\,fs) non-linear refractive index at high intensities, which  we phenomenologically described as HOKE terms from $n_4I^2$ to $n_8I^4$ in air (resp., to $n_{10}I^5$ in argon). These terms have been obtained by a numerical fit on the experimental data.
The implications to filamentation rely on two aspects of the measurement. On one hand, the fact that the Kerr effect can saturate and even become negative at high intensities is necessary to enable all-Kerr driven self-guiding as described in\,\cite{BejotKHLVHFLW2010a}. On the other hand, this result can only have practical implications on filamentation if this inversion occurs at an intensity $I_{\Delta n_{\textrm{Kerr}}=0}$ below or close to the clamping intensity predicted by the usual filamentation model relying on the balance between Kerr self-focusing and defocusing by the free electrons, i.e., $I_{\textrm{crit}}\approx$5$\times$10$^{13}$\,W/cm$^2$ in air\,\cite{KaspaSC2000,BeckeAVOBC2001}. If  $I_{\Delta n_{\textrm{Kerr}}=0}$  is higher, the Kerr inversion will not occur in laser filaments, and will therefore play a negligible role in filamentation.
We shall therefore describe in this section the experimental protocol that has been used in order to determine  the intensity and extract  the HOKE indices from the experimental sets of data,  but also discuss potential sources of artifact  affecting the measurement of these negative HOKE indices.

\subsection{Measurement  protocol}
\begin{figure}[tb!]
  \begin{center}
       \includegraphics[keepaspectratio, width=8cm]{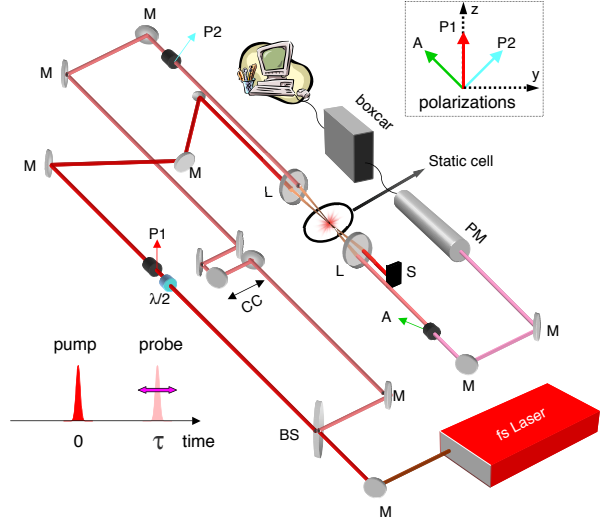}
  \end{center}
  \caption{Experimental setup for time-resolved birefringence measurements. M: Mirror, BS: Beam Splitter, L: Lens (f=20\,cm), CC: Corner Cube, P: Polarizer, A: Analyzer, S: Beam stop, PM: Photomultiplier. The relative polarizations of the pump (P1), probe (P2), and signal-field (A) are shown in the inset.}
  \label{LPhysDispoExp}
\end{figure}

In the  experiment reported in\,\cite{Loriot17_2009,Loriot18_2010}, the transient birefringence of a molecular or atomic gas sample has been measured using a time-resolved polarization technique  depicted in Fig.\,\ref{LPhysDispoExp}. The setup allows to carry out  two types of detection: homodyne and heterodyne. In the former, the signal is related  to the squared amplitude of the birefringence (amplitude-sensitive detection), whereas in the latter it is related to the birefringence itself (amplitude and phase sensitive detection). The heterodyne detection therefore provides the sign of the birefringence. In practice it is implemented by introducing between the cell and the analyzer a phase plate producing a static birefringence (see \,\cite{Loriot17_2009}). The measurements are based on the comparison between  two optical Kerr contributions induced by the field; the electronic Kerr contribution  resulting  from the deformation of the electronic cloud and the reorientation of the molecular dipole due to  molecular alignment\,\cite{Stapelfeldt75_2003}, respectively.
This section provides details about  the procedure that has been followed in order to extract the HOKE indices from this experiment. 

\subsubsection{Intensity calibration} 
\label{sec_eff_int}


A special attention  has been paid to the estimation of the laser intensity experienced by the molecules or atoms present within  the interaction volume.  This  intensity  was inferred from  the measurement of the  field-free alignment\,\cite{Seideman83_1999,Rosca87_2001}.  The last, also named post-pulse alignment,   is described by $\langle \cos^2 \theta\rangle(t)-1/3$, with $\theta$ the angle between the molecule axis and the field direction and where $\langle\,\rangle$ denotes the expectation value  averaged over the thermal distribution of molecules\,\cite{Renard70_2004}.   It is well established that both the structural shape of the alignment revivals and the permanent alignment are very sensitive to the initially applied laser intensity\,\cite{Renard90_2003}. Below saturation of the alignment, the value of $\langle \cos^2 \theta\rangle$ at the revivals of alignment increases linearly with the applied intensity\, \cite{Renard70_2004,Rouzee38_2005}, whereas between revivals (i.e.,  for permanent alignment) it grows first quadratically  and then linearly  with the intensity. Field-free alignment therefore provides an accurate and unambiguous estimation of the  laser intensity in the gas.
 \begin{figure}[tb!]
  \begin{center}
       \includegraphics[keepaspectratio, width=8cm]{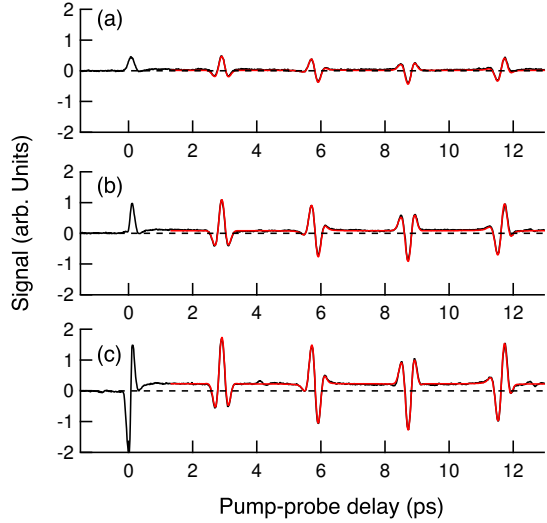}
  \end{center}
  \caption{Pump-brobe birefringence signals  (black solid lines) recorded in static cell filled with 0.1\,bar of O$_2$ at room temperature. The energy of the pump pulse is  26 (a), 56 (b) , and  96\,$\mu$J (c). Numerical fits of the postpulse signals (red solid lines) used to evaluate the  effective intensity in the experiment: $\overline{I}$=8 (a),  20 (b), and 34 TW/cm$^2$ (c), respectively (see text). The simulations are  based on Eq.\,\ref{S_cal_int}. The horizontal dashed-lines represent  the zero signal.}
  \label{fig_eff_int}
\end{figure}

Figure\,\ref{fig_eff_int} shows the time-resolved birefringence  signal of O$_2$ at different laser energies recorded with an heterodyne detection.  Pure heterodyne detection provides a post-pulse signal $S_{\textrm{hete}}$ proportional to the convolution of the probe intensity $ I_{\textrm{pr}}(t)$ with $\left(\langle \cos^2 \theta\rangle(t)-1/3\right)$ \cite{Loriot17_2009,Loriot40_2007}:
 \begin{equation}
S_{\textrm{hete}}(t)\propto I_{\textrm{pr}}(t) \otimes{\frac{3\rho\Delta\alpha}{4n_0\epsilon_0} \Big(\langle\cos^2\theta\rangle(t) -\frac{1}{3} \Big )} ,
\label{S_cal_int}
\end{equation}
with $\Delta \alpha$ the polarizability anisotropy, $\rho$  the gas density, $n_0$ the linear refractive index, $\epsilon_0$ the dielectric constant of vacuum, and $\otimes$ denotes the convolution. The permanent alignment offsets the baseline for positive delays. This offset increases  with  the intensity. Because of the intensity profile of the pump and probe beams, the signal measured in the experiment results from a spatial averaging. Using a space-averaged calculation (i.e., a 3D model) we have checked  numerically  that the volume effect can be adequately taken into account in a simpler 1D calculation just by using  an effective intensity. In fact, considering a gaussian beam profile, a peak laser intensity $I_{\mathrm{peak}}$, and a crossing angle of about $4^{\circ}$  between the pump and probe beams\,\cite{Loriot17_2009,Loriot18_2010}, the field-free alignment signal integrated over the volume can be approximated  by the  signal produced at  the effective intensity $\overline{I}$ defined as 
\begin{equation}
\overline{I}\simeq I_{\mathrm{peak}}/1.7.
\label{eff_int}
\end{equation}
 This approximation  allows to save computer time when  fitting the experimental data.

 Figure\,\ref{fig_eff_int} also shows the results of the simulations that have been used to fit the effective intensities.  The  temporal envelope $I_{\textrm{pr}}(t)$ of the probe beam  has been described by a gaussian function of duration slightly above 100\,fs (FWHM) in order to account for the crossing of the two beams. The fact that  the simulations reproduce very well both the revivals and the permanent alignment supports the analysis based on the effective intensity. 
\begin{figure}[htb!]
  \begin{center}
      \includegraphics[keepaspectratio, width=8cm]{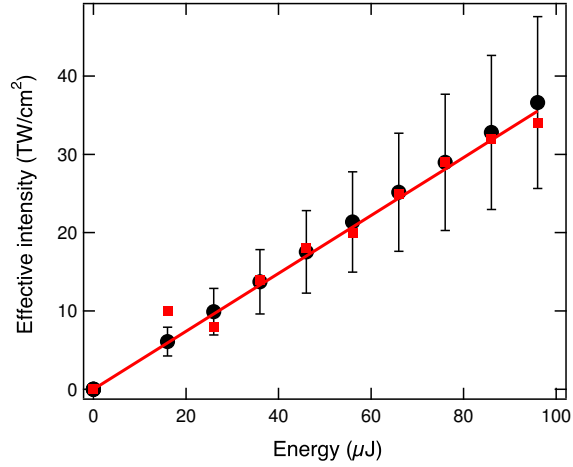}
  \end{center}
  \caption{Effective intensity $\overline{I}$ fitted on the postpulse alignment signal in O$_2$ (red solid squares) compared to the intensity estimated from the measurement of the pulse/beam characteristics (black solid circles). Linear regression of the fitted intensity on the measured pulse energy (red solid line). }
  \label{Etallonage_E_I_081013}
\end{figure}

The intensity determined as described above has been double-checked by  comparing its value  to  the one  estimated from the  measurement of the beam waist, the incident energy, and the pulse duration.  The estimation was supported by {i) the limited gas pressure  in the static cell (0.1\,bar, except for measurements at or below 1 TW/cm$^2$), {ii) the limited power of the incident beam $P = 1.8$\,GW$ \ll P_ \textrm{cr} = 80$\,GW, where $P_\textrm{cr}$ is the critical power for  nitrogen, as well as  the low value of  the nonlinear phase accumulated during propagation (less than 0.1 and 0.14 rad  for N$_2$ and O$_2$, respectively) and (iii) the focused geometry ($f$ = 20 cm), which limits self-channeling that would  induce changes in the beam profile. This  ensures that the propagation was mostly linear in the experiment, allowing to estimate the intensity therein from the measurement of the beam waist conducted at low energy. Figure\,\ref{Etallonage_E_I_081013} reports the effective intensities obtained by the two independent methods in the case of nitrogen. The error bars on the measured  intensity  (solid circles) results from the uncertainties on the measured energy ($\pm 5\,\%$), pulse duration ($\pm 10\,\%$), and beam waist ($\pm 15\,\%$). The red solid line  corresponds to a linear regression of the intensities fitted on the postpulse alignment signal.   It allows to estimate an uncertainty of  $\pm 10\,\%$ (at 3 standard deviations) in the determination of the intensities.  We can therefore exclude flaws in the intensity determination beyond this order of magnitude.

\subsubsection{Determination of the lowest-order Kerr index ($n_2$)}
The measurement of the lowest-order Kerr index  has been conducted at low energy  in order to avoid any influence from the HOKE terms. In order to ensure a good signal-to-noise ratio despite the weakness of the birefringence signal,  an homodyne detection, that offers a better sensitivity than the heterodyne one, has been employed. In the case of a  homodyne detection, the  signal is given by\,\cite{Loriot17_2009} 
\begin{eqnarray}
S_{\textrm{homo}}(t)&\propto& I_{\textrm{pr}}(t) \otimes \left( \frac{3\rho\Delta\alpha}{4n_0\epsilon_0} \Big(\langle\cos^2\theta\rangle(t) -\frac{1}{3} \Big )\right.
\nonumber \\
&&
+ \left.\frac{2}{3} n^{\textrm{cross}}_{2} I(t)\right)^2,
\label{S_homo_n2}
\end{eqnarray}
where the lowest-order electronic Kerr response $n^{\textrm{cross}}_{2}$ has been added to the retarded rotational response resulting from the alignment. Here, it is worth mentioning that the signal measured in our experiment  results from the cross-coupling between two distinguishable laser beams, namely the pump and probe beams. The coefﬁcients $n^{\textrm{cross}}_{2}=2\times n_2$, with $n_2$ the standard self-induced Kerr index, therefore describes the non-linear refractive  index due to cross-Kerr effect. The correspondence  between cross-Kerr and Kerr  indices is given  in\,\cite{Loriot18_2010}. Finally, the factor 2/3 results  from  the different  values of the Kerr index   experienced by the probe field along directions parallel ($\parallel$) and perpendicular ($\perp$) to the pump field, respectively,  with $n_{2_{\parallel}}=3\times n_{2_{\perp}}$. This relation  is valid when the intrapulse alignment can be neglected so that the medium can be viewed as isotropic during the pump excitation\,\cite{Boyd_2007}.  The approximation  is justified for the investigated molecules and the  relatively short pulse duration used in the experiment compared to the rotational period.  For instance, in N$_2$ or O$_2$, the orientational Kerr contribution to $n_2$ calculated  from  the elements of the hyper-polarizability tensor\,\cite{Lalanne1996}  is less than  5\% at the maximum peak intensity investigated in the experiment.
\begin{figure}[htb!]
  \begin{center}
      \includegraphics[keepaspectratio, width=8cm]{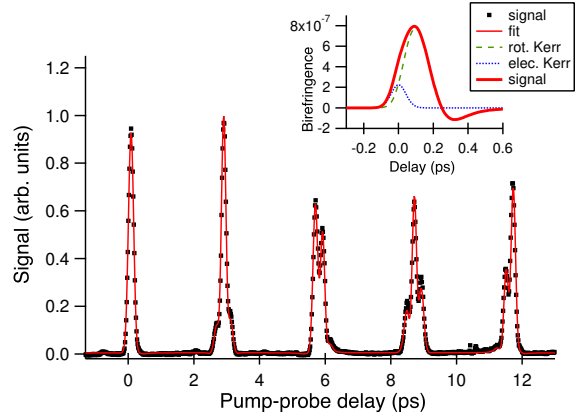}
  \end{center}
  \caption{Homodyne birefringence signal (black solid squares) recorded in 1~bar of O$_2$ at room temperature and numerical fit  (red  solid line). The effective intensity of the pump is 1\,TW/cm$^2$. In the inset, the total birefringence  (red solid line) resulting  from the electronic (blue dotted line) and rotational (green dashed line) Kerr contributions.}
  \label{EffetKerr_nN2_n2O2}
\end{figure}  

Since the permanent alignment can be neglected  in the low  intensity regime, both rotational and electronic contributions exhibit a linear dependency  on the applied  pump intensity $I$, which therefore acts as an amplitude factor  on the overall signal. $n^{\textrm{cross}}_{2}$ was hence determined independently from the knowledge of this intensity. Figure\,\ref{EffetKerr_nN2_n2O2} shows the time-resolved birefringence signal recorded in O$_2$ at low intensity. For all recorded data, the value of $n_2$ has been evaluated through  deconvolution of the rotational response from the signal.  First, the postpulse signal  was  adjusted by  Eq.\,\ref{S_homo_n2} with $n^{\textrm{cross}}_{2}=0$, using a low intensity value as a fixed parameter and an amplitude factor  as a free parameter. This allows to evaluate the contribution of the rotational term close to the  zero delay, and only relies on the knowledge of the molecular polarisability ($\Delta\alpha=$ 4.6   and 7.25 a.u. for N$_2$ and O$_2$, respectively). Next, the signal  was adjusted  using  $n^{\textrm{cross}}_{2}$ as a  free parameter.   As described in \cite{Loriot17_2009,Loriot18_2010}, measurements in argon have been calibrated with the postpulse signal recorded independently in N$_2$ in  the same experimental conditions.  The data presented in Table\,1 of \cite{Loriot17_2009,Loriot18_2010} result from statistics performed over 50, 30, and 8  data samples recorded at low intensity (i.e., $\overline{I}<1$\,TW/cm$^2$) in N$_2$, O$_2$, and Ar, respectively.

\subsubsection{Determination of the higher-order Kerr indices}
\label{HOKsection}
All HOKE indices have been determined by the use of an heterodyne detection that is phase sensitive and therefore allows to recover  the sign of each term. When considering HOKE terms up to the fifth power of the applied intensity $I$, the heterodyne birefringence  signal is given by\,\cite{Loriot17_2009}
\begin{eqnarray}
S_{\textrm{hete}}(t)&\propto& I_{\textrm{pr}}(t) \otimes \rule[-3 mm]{0 mm}{3 mm} \left(\Delta n_\textrm{rot}(t)+\frac{2}{3} n^{\textrm{cross}}_2 I(t)\right.
\nonumber \\
&&
+\frac{4}{5} n^{\textrm{cross}}_4 I(t)^2 +\frac{6}{7} n^{\textrm{cross}}_6 I(t)^3
\nonumber \\
&&
+\left.\frac{8}{9} n^{\textrm{cross}}_8 I(t)^4+\frac{10}{11} n^{\textrm{cross}}_{10} I(t)^5\rule[-3 mm]{0 mm}{3 mm}\right),
  \label{S_hete}
\end{eqnarray}
with $\Delta n_\textrm{rot}(t)= 3\rho\Delta\alpha/4n_0\epsilon_0 \left(\langle\cos^2\theta\rangle(t) -\frac{1}{3} \right )$. We have generalized  the relation $n_{2j_{\parallel}}=(2j+1)\times n_{2j_{\perp}}$  with $j\in \mathbb{N}^{*}$, verified for $n_{2}$ and $n_{4}$\,\cite{Arabat10_1993}, to higher orders. As mentioned in the previous section, this approximation results from neglecting the intrapulse alignment considering  the medium as isotropic during the interaction with the pump.

Although the determination of $n^{\textrm{cross}}_2$ is straightforward, since it is independent from $I$, the evaluation of the HOKE indices is complicated by the fact that spatial averaging depends on the non-linearity order. As in Sec.\,\ref{sec_eff_int}, in order to  avoid the prohibitively large use of computer time, the deconvolution of the HOKE  from the signal has been achieved by fitting the experimental data with a 1D model. However,  in order to evaluate the influence of the volume effect,  different 3D simulations of Eq.\,\ref{S_hete} have been preliminary performed. First, Eq.\,\ref{S_hete}  has been truncated to the second power  of $I$ (i.e., $n^{\textrm{cross}}_6, \cdots, n^{\textrm{cross}}_{10}=0$) and  then  spatially integrated. Second, the 1D model, in which $n^{\textrm{cross}}_4$ was  replaced  beforehand by the effective index $\overline{n}^{\textrm{cross}}_4$, has been used to fit the previous numerical result using the  effective intensity defined in Sec.\,\ref{sec_eff_int} as a fixed  parameter and  $\overline{n}^{\textrm{cross}}_4$ as a free parameter. For the next HOKE index $n^{\textrm{cross}}_6$, the same procedure has been applied. The result of the 3D calculation including $n^{\textrm{cross}}_4$ and $n^{\textrm{cross}}_6$ has been fitted  with the 1D model with $\overline{n}^{\textrm{cross}}_4$ fixed,  $\overline{n}^{\textrm{cross}}_6$ being the free parameter. This approach has been repeated successively for  each HOKE index up to $n^{\textrm{cross}}_{10}$. In order to check that the ratios between the HOKE indices and their respective effective values were independent from the intensity, different numerical tests have been performed over the intensity range considered experimentally.  The ratio between the HOKE indices and the effective values that account for the volume effect are given in Table\,\ref{table_ni_eff}. Since both $n_2$ and the alignment depend linearly on the intensity, the correction factor for $n_2$ is $c_2=1$ (i.e., $\overline{n}^{\textrm{cross}}_2=n^{\textrm{cross}}_2$).
\begin{table}[tdp]
\begin{center}
\begin{tabular}{|c|c|c|c|c|}
$n^{\textrm{cross}}_{2j}$ & $n^{\textrm{cross}}_4$   & $n^{\textrm{cross}}_6$ &$n^{\textrm{cross}}_8$ & $n^{\textrm{cross}}_{10}$\\
$c_{2j}$ &  $1.04$ & $1.30$  &$1.45$ & 2.66
\end{tabular}
\end{center}
\caption{Correction factors for the volume effect defined as $c_{2j}=\overline{n}^{\textrm{cross}}_{2j}/n^{\textrm{cross}}_{2j}$, with $n^{\textrm{cross}}_{2j}$ and $\overline{n}^{\textrm{cross}}_{2j}$  the HOKE terms and its corresponding  effective value, respectively.}
\label{table_ni_eff}
\end{table}
\begin{figure}[htb!]
  \begin{center}
      \includegraphics[keepaspectratio, width=8.5cm]{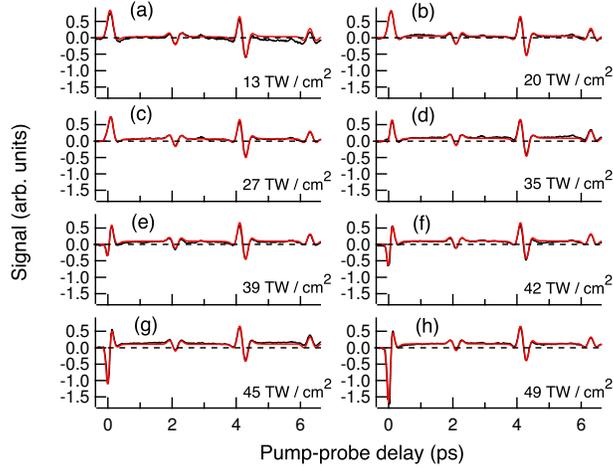}
  \end{center}
  \caption{Intensity dependence of the  birefringence signals  (black solid lines) recorded in static cell filled with 0.1\,bar of N$_2$ at room temperature. Numerical simulations of Eq.\,\ref{S_hete} (red solid lines) where $n^{\textrm{cross}}_{2j}$ and $I$ have been replaced by the effective values $\overline{n}^{\textrm{cross}}_{2j}$ and $\overline{I}$, respectively (see text).  $\overline{I}$=13 (a), 20 (b),  27 (c),  35  (d),  39 (e), 42 (f),  45 (g), and 49\,TW/cm$^2$ (h).}
  \label{N2_High_int}
\end{figure} 

Figure\,\ref{N2_High_int} displays the time-resolved birefringence signals recorded in N$_2$ from low to high intensity. The simulations have been performed using the $n^{\textrm{cross}}_{2j}$ coefficients first reported in\,\cite{Loriot17_2009,Loriot18_2010}, determined as detailed above. Like in O$_2$ and Ar,  the electronic Kerr contribution to the birefringence signal (i) scales  linearly  with the intensity at low intensity where $n^{\textrm{cross}}_2$ dominates, (ii) then saturates at moderate intensity, and finally (iii) reverses its sign and becomes highly non-linear at high intensity where the HOKE dominates. To avoid correlation between the $n^{\textrm{cross}}_{2j}$ coefficients in the least square fitting procedure orders were fitted successfully. $\overline{n}^{\textrm{cross}}_4$ was first adjusted with $n^{\textrm{cross}}_2$ fixed and $\overline{n}^{\textrm{cross}}_{2j>4}=0$, from a set of 14 (resp., 10 and 11) data  recorded in N$_2$ (resp., O$_2$ and Ar)  at an effective intensity $\overline{I}\leqslant$27\,TW/cm$^2$ (resp., 20 and 24\,TW/cm$^2$). Due to the predominant value of   $\overline{n}^{\textrm{cross}}_8$ compared to   $\overline{n}^{\textrm{cross}}_6$, it has not been possible to isolate an intensity window where these two indices could be fitted independently.  They have therefore  been determined simultaneously, with $n^{\textrm{cross}}_2$ and  $\overline{n}^{\textrm{cross}}_4$ fixed and $\overline{n}^{\textrm{cross}}_{10}=0$, from a set of  24 (resp., 10 and 26) data recorded  in N$_2$ (resp., O$_2$ and Ar)  at $27<\overline{I }<50$\,TW/cm$^2$ (resp., $20<\overline{I}<35$\,TW/cm$^2$ and $24<\overline{I}<45$\,TW/cm$^2$). For the same reason, the last term $\overline{n}^{\textrm{cross}}_{10}$,  necessary only  in argon, has been fitted together with $\overline{n}^{\textrm{cross}}_6$ and $\overline{n}^{\textrm{cross}}_8$.

In order to reduce the errors bars, the records mentioned above  have also been used in order to fit the HOKE indices on the two-dimensional data displaying birefringence signal as a function of intensity and time delay.
 Figure\,\ref{Ajust2D_N2_O2_Ar} displays a comparison between the experimental data set and the numerical simulations resulting form the fitting procedure. The good agreement between observations and calculations confirm the values of the HOKE indices retrieved as detailed above, as evidenced in particular by the fact that they overcome the rotational response (see positive signal at positive delays for N$_2$ and O$_2$) at large intensity.
 \begin{figure}[htb!]
  \begin{center}
      \includegraphics[keepaspectratio, width=9cm]{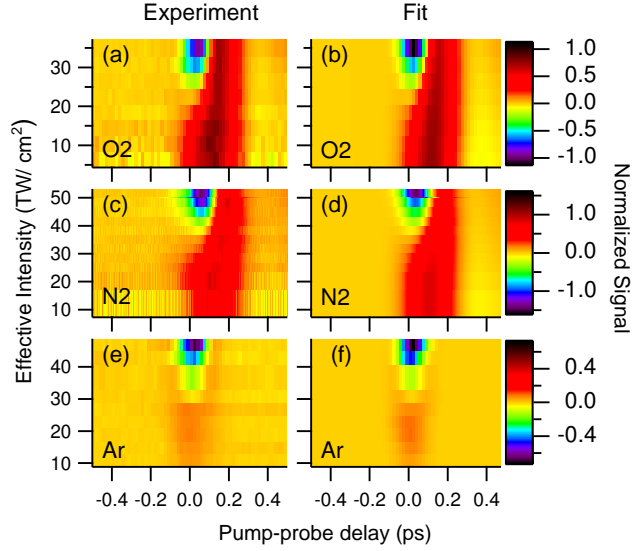}
  \end{center}
  \caption{Time-resolved birefringence signals recorded for different effective intensities  $\overline{I}$ in O$_2$ (a), N$_2$ (c),   and Ar (e). Numerical fits of the  O$_2$ (b), N$_2$ (d),   and Ar (f)  data normalized for each intensity to the maximum amplitude of the postpulse  signal (not shown), except for argon.}
  \label{Ajust2D_N2_O2_Ar}
\end{figure} 

\subsection{Is the observed inversion due to negative HOKE?}
Negative non-linear indices generated by high-power laser pulses are generally attributed to a contribution of the free electrons, given by the Drude model \cite{BergeSNKW2007}:
\begin{equation}
\Delta n_{plasma}=-\frac{\rho}{2n_0\rho_{crit}},
\end{equation}
where $\rho$ is the electronic density, $n_0$ is the linear refractive index, $\rho_{crit}=\frac{\epsilon_0 m_e \omega_l^2}{e^2}$ is the critical plasma density ($\rho_{crit}$ = 1.75$\times$10$^{27}$ m$^{-3}$ at 800 nm), $\epsilon_0$ is the permittivity of vacuum, $m_e$ is the electron mass, $\omega_l$ is the laser angular frequency, and $e$ the elementary charge.

As described in the previous section, our measurements were performed in a pump-probe configuration where  the pump beam intensity is sufficient to partially ionize the gas, so that the free electrons contribute negatively to the refractive index. To cancel this free electron contribution, we measured the transient variation of the \emph{birefringence} rather than the variation of the refractive index itself. This approach is supported by the general belief that non relativistic plasmas are not birefringent, even at the time scale of the laser-medium interaction \cite{Boyd_2007}. However, since this statement has never been proved rigourously to date, we discuss below the potential plasma contribution on each time scale.

\subsubsection{Postpulse plasma contribution} 
The birefringence induced by a high-intensity ultrashort pulse in argon is not maintained after the pulse has passed \cite{B'ejPBKMW2008a,PetitBBKMW2009,MarceRGWCTCDSC2010}. A direct contribution of the plasma to the medium birefringence  would have the same lifetime as the plasma, i.e., at least several ps \cite{TzortPFM2000}. It can therefore be excluded.
 \begin{figure}[htb!]
  \begin{center}
      \includegraphics[keepaspectratio, width=8cm]{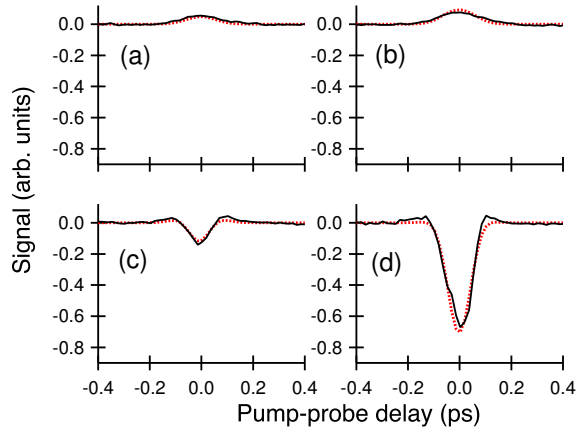}
  \end{center}
  \caption{Time-resolved birefringence signals (solid lines) recorded for different effective intensities  $\overline{I}$=10 (a), 21 (b), 28 (c), and 43 TW/cm$^2$ (d) in argon. Numerical fits are displayed as red dotted lines.}
  \label{Graphajust_normaliz_2}
\end{figure}

This negligible birefringence of the plasma is further confirmed by the temporal dependence of the birefringence in our experiments. In the case of argon (see Fig.\,\ref{Graphajust_normaliz_2}) the birefringence goes down to zero for time delays larger than $\sim$200\,fs, i.e., as soon as the two pulses do not overlap anymore. In the case of oxygen and nitrogen (see Figs.\,\ref{fig_eff_int} and \ref{N2_High_int}, respectively), the birefringence observed  at positive delays between the revivals of alignment is perfectly reproduced by the molecular permanent alignment described in Sect.\,\ref{sec_eff_int}.  

\subsubsection{Intrapulse plasma contribution} 
Non-birefringent plasma could however be expected to reduce the birefringence by inducing an equal refractive index change on both axes. However, free electrons accumulate over the pulse duration, so that this potential contribution to the non-linear birefringence should be asymmetric in time, as depicted in Fig.\,\ref{asymetrie}. In contrast, for a symmetric pulse like we used in\,\cite{Loriot17_2009}, the Kerr contribution to the refractive index is symmetrical in time. The relative contribution of the plasma to the birefringence would be therefore defined by the asymmetry of the birefringence profile around time $t=0$. Such  asymmetry is not observed in our experiment, as  evidenced for example by the data in argon displayed in Fig.\,\ref{Graphajust_normaliz_2}.
\begin{figure}[tb!]
  \begin{center}
      \includegraphics[keepaspectratio, width=8cm]{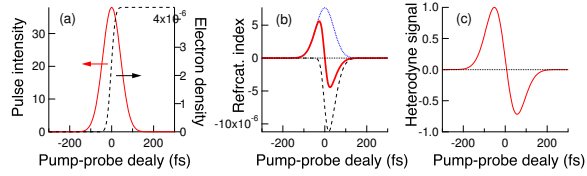}
  \end{center}
  \caption{(a) Pulse intensity (red line) and electron density (black dotted line) calculated in argon along the pulse duration (100\,fs) for a peak intensity of 28\,TW/cm$^2$. (b) Corresponding refractive index produced by  a positive Kerr effect (blue dotted line), an hypothetical  negative Kerr effect resulting from  accumulated free electrons (black dashed line), and both contributions (red solid line). (c) Heterodyned signal resulting from both contributions.}
  \label{asymetrie}
\end{figure}

Besides the free electrons, the plasma is made of ions. Their susceptibility can be estimated according to Sprangle \textit{et al.} \cite{SpranEH1997}:
\begin{equation}
\frac{\chi^{(3)}_{\textrm{ion}}}{\chi^{(3)}_{\textrm{neutral}}}\approx\left(\frac{\textrm{IP}_{\textrm{neutral}}}{\textrm{IP}_{\textrm{ion}}}\right)^3,
\end{equation}
where $\textrm{IP}_x$ denotes the ionization potential of species $x$, as detailed in Table \ref{IP}. Assuming that the refractive index variation of the ions is three times larger along the laser polarization axis than perpendicular to it, as is the case for the neutrals, the contribution of the ions is about 10\% to 20\% lower than that of neutrals for the same partial pressure. At the maximum intensity used our experiment, the ionization is restricted to 1\% at most\,\cite{Talebpour163_1999}, so that the ions cannot, in any case, contribute to more than 0.2\% to the observed birefringence. Moreover, as discussed above for the electrons, any contribution from the ions would result  in an asymmetric temporal  profile.
\begin{table}[tdp]
\begin{center}
\begin{tabular}{|c|c|c|c|}
Species & N$_2$ & O$_2$ & Ar \\
IP$_{\textrm{neutral}}$(eV) & 15.6 & 12.07 & 15.8 \\
IP$_{\textrm{ion}}$(eV) & 27.12 & 24.14 & 27.6 \\
$\frac{\chi^{(3)}_{\textrm{ion}}}{\chi^{(3)}_{\textrm{neutral}}}$ & 0.19 & 0.125 & 0.185
\end{tabular}
\end{center}
\caption{Ionization potentials of neutral and ionized nitrogen, oxygen, and argon.}
\label{IP}
\end{table}

Finally, ionization can deplete the ground state population to the benefit of excited bound and continuum states. Both processes can result in a large modification of the refractive index\,\cite{Nurhuda66_2002,Nurhuda10_2008,SteinBD2010}. 
Moreover, for intensities close to the inversion of the refractive index, the ionization takes place at the frontier between multiphoton and tunneling ionization regimes. Besides depletion of the ground state, ionization might contribute to the HOKE through the fast-moving electrons recolliding with the atomic or molecular core, as described by  the  three-step model\,\cite{Corkum13_1994}. The resulting oscillation of the induced dipole responsible for high harmonic generation\,\cite{Itatani432_2004} could lead to large nonlinearities  in the Kerr effect.  Since we consider a process occurring within an optical cycle, this potential effect can be considered as instantaneous as compared with the time scale of the pulse. Calculations of the  dipole induced by a strong laser field could therefore contribute to interpret the effect observed  in our experiment.

\subsubsection{Two-beam coupling}
Another artifact that could be raised  is the energy exchange between the  two crossing laser pulses, known as two-beam coupling\,\cite{Boyd_2007,Dogariu14_1997}. First, we  should mention that the signal provided by heterodyne detection used in the high-order Kerr measurement  is  in principle  free from any  two-beam coupling contribution. In fact, pure heterodyne detection results from the difference between two data set recorded in the same conditions, except for opposite phases of a local oscillator\,\cite{Minhaeng99_1993,Lavorel31_2000}.  Second, for femtosecond pulses with identical spectra, two-beam coupling requires a frequency chirp and a  finite time response of the nonlinear refractive index \,\cite{Smolorz17_2000}. The very fast excitation time  associated with non-resonant excitation combined with  the small residual frequency chirp of our pulses can only lead to a marginal amount of two-beam coupling through the purely electronic Kerr response. Finally, and most importantly, the time profile resulting from two-beam coupling should be asymmetric with respect to the pump-probe delay. This was not observed in our experiment, as previously mentioned (See e.g. Fig. \ref{Graphajust_normaliz_2}). Consequently, the influence of  two-beam coupling on our experimental results can be confidently disregarded.

\section{HOKE- and plasma-driven filamentation regimes}

Improving models by considering higher-order of the relevant processes is a quite natural approach in all branches of physics. Higher-order optical non-linearity is well-known, giving rise e.g. to higher harmonics \cite{Boyd_2007} and should therefore be included in the propagation equations, unless the it is proven that it induces a negligible effect in the considered situation. In the context of filamentation, such development was already tried several years ago \cite{AkoSBC2001,VincoB2004,FibicI2004,BergSMKYFSW2005,ZhangTWDW2010}, although the significance of those works was limited by the lack of experimental knowledge of the magnitude of these terms.

This magnitude constitutes the key question arising about their role in laser filamentation. Considering the wide range of wavelengths, pulse durations, incident energies and focusings
investigated in the last ten years \cite{ChinHLLTABKKS2005,BergeSNKW2007,CouaiM2007,KaspaW2008}, we expect that no unique dominant effect can be identified for all conditions. Rather, four regularizing terms exist in the full non-linear Schrödinger equation (NLSE) \cite{BejotKHLVHFLW2010a}, with relative contributions depending on the experimental conditions. Three of them (defocusing by the HOKE and by the free electrons, as well as diffraction) are spatial, while group-velocity dispersion (GVD) is temporal.

The latter two are independent on intensity. Diffraction cannot be the dominant regularizing factor in filamentation, since it can only occur beyond the self-focusing $P_\textrm{c}$, defined as the power required for the self-focusing to dominate diffraction. Furthermore, due to the limited bandwidths at play and the relatively low dispersion of usual gases in the near infrared, the contribution of GVD is very small in usual filamenting conditions, especially on the short distances of the laboratory. However, for few-cycle pulses centered in the UV, the bandwidth and the dispersion increase, so that GVD may play a substantial role as it does e.g. in water \cite{DubieGTT2004}.

In most usual cases, defocusing by plasma and/or the HOKE will therefore be the main regularizing terms. The relative contributions of the HOKE and the plasma-induced refractive index change can be characterized as the ratio of the corresponding terms in the NLSE, at any location $\vec{r}$ and time $t$ \cite{EttouBPLHFLKW2010}:
\begin{equation}
\xi(\vec{r},t)= \lvert \sum_{j\ge2}{n_{2j} I(\vec{r},t)^j}\rvert / \frac{\rho(\vec{r},t)}{2 n_0\rho_{\textrm{crit}}}.
\label{xi}
\end{equation}

The overall action of both effects on the whole pulse duration is described by the pulse-integrated value of $\xi$:
\begin{equation}
\overline{\xi}(\vec{r})= \frac{ \int \lvert \sum_{j\ge2}{n_{2j} I(\vec{r},t)^j}\rvert \times |\varepsilon(\vec{r},t)| \mathrm{d}t} 
 {\int \frac{\rho(\vec{r},t)}{2 n_0\rho_{\textrm{crit}}} \: |\varepsilon(\vec{r},t)| \mathrm{d}t},
\end{equation}
where $\varepsilon$ is the reduced scalar envelope defined such that $|\varepsilon|^2=I$ \cite{EttouBPLHFLKW2010}.

Both HOKE and plasma defocusing generate a negative non-linear refractive index. However, they differ by three aspects. First, multiphoton ionization requires 8 photons in O$_2$ and 11 photons in N$_2$ and Ar, so that the corresponding non-linearity is of 8th- or 11th order. In contrast, over the intensity range relevant for filamentation, orders up to 4 (resp. 5) only have to be considered in air (resp. argon), as discussed in Sec.\,\ref{HOKsection}. Secondly, the Kerr effect has a time constant of a few fs at most, shorter than the pulse duration, while the plasma density accumulates over whole duration of the same pulse. Finally, both terms increase in absolute values for higher frequencies, but the Kerr effect shows a relatively slow dispersion over the visible spectral range \cite{EttouPKW2010}, while the increase of the ionization rates is much faster \cite{BergeSNKW2007}, as illustrated in Fig.\,\ref{dispersion_kerr_plasma} in the case of air. 
\begin{figure}[tb!]
  \begin{center}
      \includegraphics[keepaspectratio, width=8.2cm]{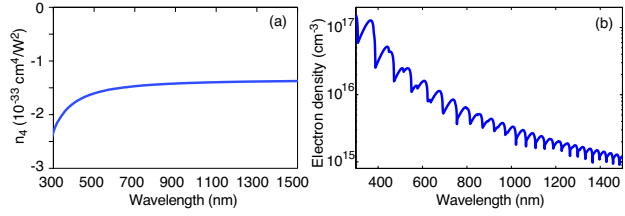}
  \end{center}
  \caption{(a) Dispersion of the fifth-order Kerr index ($n_4$) \cite{EttouPKW2010} and (b) electron density generated in air by a 30 fs pulse of constant intensity 50~TW/cm$^2$. Note that the vertical scales are respectively linear and logarithmic.}
  \label{dispersion_kerr_plasma}
\end{figure}

One can therefore expect that the plasma will tend to provide the dominant defocusing contribution on the short-wavelength side, while the HOKE should be favoured on the long-wavelength side. Indeed, we observed this transition in a recent numerical work \cite{EttouBPLHFLKW2010}. As a consequence, one can consider as a general rule that the regularizing process will be defocusing by the free electrons for short wavelengths ($\lambda \le$ 300-400 nm) and by the HOKE for longer wavelengths.
The discussion should therefore focus on defining the border between ``long'' and ``short" wavelengths, rather than on the existence of the HOKE. Still, the position of this transition depends little on the value of the HOKE indices since the ionization coefficients cover several orders of magnitudes over the spectrum. 
\begin{figure}[tp!]
  \begin{center}
      \includegraphics[keepaspectratio, width=8.4cm]{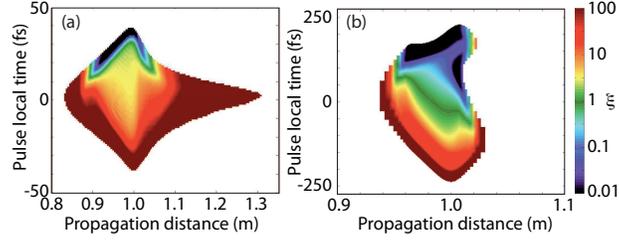}
  \end{center}
   \caption{Numerical simulation of the space-time dynamics of the on-axis ratio $\xi(r=0)$ of the respective contributions of higher-order Kerr terms and plasma to defocusing (see text for detail). The white regions correspond to irrelevant values due to the low intensity ($I<I_{\textrm{max}}/e^2$). Pulse durations are (a) 30 fs ($\overline{\xi}_{\textrm{min}}(r=0) = 3.3$ at $z$ = 98.7 cm) and (b) 250 fs ($\overline{\xi}_{\textrm{min}}(r=0) = 0.5$ at $z$ = 99.8 cm). Note the asymmetric temporal shape due to the plasma accumulation, contrasting with the symmetric pulse shape related with the domination of the instantaneous HOKE in panel (a).}
  \label{temporel}
\end{figure}

Similarly, as a consequence of the different temporal dynamics of the plasma and the HOKE, the pulse duration will strongly impact their respective contribution to defocusing. The longer the pulse, the more efficient the accumulation of plasma, and consequently its relative contribution to defocusing. Indeed, as shown in Fig.\,\ref{temporel}a, numerical calculations using the
model described in \cite{EttouBPLHFLKW2010} and relying on the generalized non-linear Schr\"odinger equation and 
 ionization rates given by the multi-species generalized Keldysh-Perelomov, Popov, Terent'ev (PPT) formulation \cite{BergeSNKW2007}, the plasma contribution is marginal ($\xi \gg 1$) over most of the duration of a short pulse. It only plays a significant role at the very tail of the pulse, where the intensity has decreased close to zero. Its contribution to the overall pulse propagation is therefore negligible, as evidenced by a high on-axis value of $\overline{\xi}_{min}(r=0) = 3.3$ for a 30 fs pulse. In contrast, for a longer pulse (250 fs, Fig.\,\ref{temporel}b), the plasma efficiently accumulates earlier in the pulse and is significant already in its high-intensity region. It therefore contributes significantly ($\overline{\xi}_{min}(r=0) = 0.5$ for a 250 fs pulse) to the propagation of long pulses, resulting in particular in an asymmetric temporal pulse shape. The published values of the HOKE indices suggest that the border between the two regimes lies at a few hundreds of fs at 800 nm. The two regimes are therefore experimentally accessible with the current laser technology, which may explain the contradicting results obtained among the available experimental work. Furthermore, the transition between them is smooth, so that in adequate conditions both processes contribute with similar magnitudes.
Again, the discussion should focus on the domains of the long- and short pulse regime rather than on seeking for one single universally dominating mechanism.

The different regimes discussed above can be displayed schematically as a ``phase diagram" (Fig.\,\ref{diagramme_phase}). In this graph, the axis have been left blank on purpose to ensure the generality, independently of the measurement of the HOKE terms. From a practical point of view, the transition between the regimes corresponds to the equality of the contributions of plasma and HOKE to defocusing ($\overline{\xi}=1$).
 The historical trend over the last 10 years includes a substantial shortening of the pulses, from 100-200 fs in the late 1990's and early 2000's to $\sim$30 fs nowadays. According to our results, this shortening corresponds to a transition between ``long" and `short" pulses, i.e., from plasma- to HOKE-regularized filaments, which might explain why HOKE had been considered marginal up to now.
 \begin{figure}[tp!]
  \begin{center}
      \includegraphics[keepaspectratio, width=8cm]{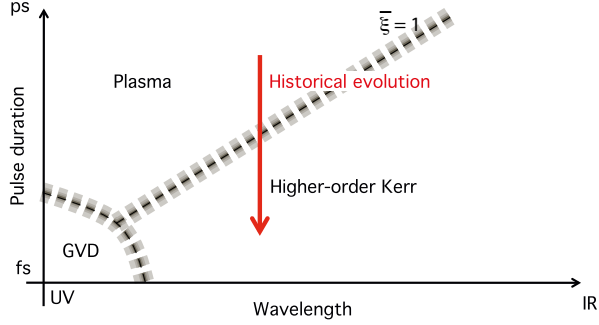}
  \end{center}
  \caption{Schematic "Phase diagram" of the dominating mechanisms at play in filamentation}
  \label{diagramme_phase}
\end{figure}

\section{Conclusion}

As a conclusion, we have discussed the details of the recent measurement of the HOKE indices \cite{Loriot17_2009,Loriot18_2010} and excluded a range of potential artifacts, confirming the reliability of these experimental data. Furthermore, we have shown that, due to different temporal and spectral dynamics of the plasma- and HOKE- induced defocusing, the former should be dominant for long pulses at short-wavelengths, while the latter will dominate for short pulses at long wavelengths. We therefore suggest that the controversy about the mechanism of filamentation should turn from a qualitative discussion of which effect dominates, to a more quantitative discussion about the domains of relevance of each process.

Acknowledgements. This work was supported by the Conseil R\'egional
de Bourgogne, the ANR \textit{COMOC}, the \textit{FASTQUAST} ITN Program of the 7th FP and the Swiss NSF (contracts 200021-125315 and 200021-125315-2).

\bibliographystyle{jeos}


\end{document}